\documentstyle[preprint,aps,prb,amsfonts]{revtex}

\newtheorem{theorem}{Theorem}
\newtheorem{corollary}[theorem]{Corollary}

\newtheorem{definition}[theorem]{Definition}

\def\stackunder#1#2{\mathrel{\mathop{#2}\limits_{#1}}}%
\def\limfunc#1{\mathop{\rm #1}}%
\def\proof{Proof. }
\def\endproof{\mbox{\ \rule{.1in}{.1in}}}

\begin{document}
\title{A ${\Bbb Z}_3$-graded generalization of supermatrices}
\author{Bertrand Le Roy\\
Laboratoire de Gravitation et Cosmologie Relativistes\\
Universit\'e Pierre et Marie Curie - CNRS URA 769\\
Tour {\bf 22}-12, 4$^e$ \'etage, bo\^\i te 142}
\maketitle

\begin{abstract}
We introduce ${\Bbb Z}_3$-graded objects which are the generalization of the
more familiar ${\Bbb Z}_2$-graded objects that are used in supersymmetric
theories and in many models of non-commutative geometry. First, we introduce
the ${\Bbb Z}_3${\bf -graded Grassmann algebra}, and we use this object to
construct the ${\Bbb Z}_3${\bf -matrices}, which are the generalizations of
the {\bf supermatrices}. Then, we generalize the concepts of {\bf supertrace}
and {\bf superdeterminant}.
\end{abstract}
\newpage

\section{Introduction}

${\Bbb Z}_2$-graded algebras, which are the basic objects of supersymmetry,
are well known since the works of F.A.~Berezin, G.I.~Kac, D.A.~Leites,
J.~Wess and B.~Zumino \cite
{berezin1970,berezin1975,wess1974,wess1977,wess1978} who introduced the
concepts of supermatrices, supertrace and superdeterminant. These concepts
will be generalized here.

Recently, there have been many attempts to generalize ${\Bbb Z}_2$-graded
constructions to the ${\Bbb Z}_3$-graded case\cite
{kerner1991,kerner1993,kerner1995,chung1993,mohammedi1994}. Many such
attempts, though, were aimed at the description of exotic statistics. We
think that our construction could describe some properties of the {\em quarks%
}, in particular the {\em ternary} aspects of their associations.

\section{The ${\Bbb Z}_3$-graded Grassmann algebra}

The ordinary Grassmann algebra is generated by anticommuting entities: 
\[
\eta _i\eta _j=-\eta _j\eta _i 
\]

This can be viewed in the following way. ${\Bbb Z}_2$ is the group of the
permutations of two elements. It can be represented by the real numbers $%
(-1) $ and~$1$: 
\[
\begin{array}{cc}
\left( 
\begin{array}{ll}
A & B \\ 
A & B
\end{array}
\right) & \left( 
\begin{array}{ll}
A & B \\ 
B & A
\end{array}
\right) \\ 
+1 & -1
\end{array}
\]

When an element of the permutation group is applied to the product of two
generators, it is followed by the multiplication by the number representing
this permutation.

Similar operations can be performed with ${\Bbb Z}_3$, which is faithfully
represented by the complex numbers $1$, $j$ and $j^2$, where $j$ is a cubic
root of $1$, $e^{\frac{2i\pi }3}$: 
\[
\begin{array}{ccc}
\left( 
\begin{array}{lll}
A & B & C \\ 
A & B & C
\end{array}
\right) & \left( 
\begin{array}{lll}
A & B & C \\ 
B & C & A
\end{array}
\right) & \left( 
\begin{array}{lll}
A & B & C \\ 
C & A & B
\end{array}
\right) \\ 
1 & j & j^2
\end{array}
\]

When one applies an element of the cyclic permutation group to the product
of {\em three} generators, it is multiplied by the complex number
representing this permutation: 
\[
\theta _i\theta _k\theta _l=j\theta _k\theta _l\theta _i 
\]

The generators defined in this way have the following properties: their
cubes vanish, as also does a product of any four generators\cite{kerner1995}%
. 
\[
\theta _i^3=0 
\]
\[
\theta _i\theta _k\theta _l\theta _m=0 
\]

Therefore the dimension of the algebra generated by $N$ independant elements
is 
\[
\stackunder{\theta ^{\prime }s}{\underbrace{N}}+\stackunder{\theta \theta
^{\prime }s}{\underbrace{N^2}}+\stackunder{\theta \theta \theta ^{\prime }s}{%
\underbrace{\frac{N^3-N}3}} 
\]

To restore the symmetry between grades 1 (the $\theta $'s) and 2 (the $%
\theta \theta $'s), we can add $N$ grade 2 generators $\overline{\theta }_i$
that behave like the products of two $\theta $'s, that is 
\[
\overline{\theta }_i\overline{\theta }_k\overline{\theta }_l=j^2\bar \theta
_k\bar \theta _l\bar \theta _i 
\]
\[
\theta _i\bar \theta _k=j\bar \theta _k\theta _i 
\]

In the case of the ordinary Grassmann algebra, the products of an odd number
of generators are automatically anticommutative, whereas the products of an
even number of generators commute with all other elements. In the ${\Bbb Z}%
_3 $-graded case, this is no longer true. For example, 
\[
\theta _i\underbrace{\theta _k\bar \theta _l}=j\underbrace{\theta _i\bar
\theta _l}\theta _k=j^2\bar \theta _l\theta _i\theta _k\mbox{;} 
\]
but in the same time, $\theta _k\bar \theta _l$ and $\bar \theta _l\theta _i$%
, as grade 0 elements, should commute with all other elements, leading to
the relations $\theta _i(\theta _k\bar \theta _l)=(\theta _k\bar \theta
_l)\theta _i$ and $\theta _i\theta _k\bar \theta _l=j^2(\bar \theta _l\theta
_i)\theta _k=j^2\theta _k(\bar \theta _l\theta _i)$, which are clearly
contradictory.

This leads us to impose the following relations on all elements of a
definite grade (the grade of the product of two elements being the sum of
their grades, modulo 3). Let us denote by $a,b,\ldots $ the elements of
grade $0$, by $A,B,\ldots $ the elements of grade $1$, by $\bar A,\bar
B,\ldots $ the elements of grade $2$ and by ${\cal X}$ an element of
arbitrary grade. Then the rules 
\[
\left\{ 
\begin{array}{rcl}
a{\cal X} & = & {\cal X}a \\ 
A\bar A & = & j\bar AA
\end{array}
\right. 
\]
define entirely the ${\Bbb Z}_3$-graded Grassmann algebra. We obtain the
ternary rules $\theta _i\theta _k\theta _l=j\theta _k\theta _l\theta _i$ and 
$\overline{\theta }_i\overline{\theta }_k\overline{\theta }_l=j^2\bar \theta
_k\bar \theta _l\bar \theta _i$ directly from the second rule by replacing $%
A $ and $\bar A$ with products of one or two generators $\theta $ or $\bar
\theta $.

With the unit element ${\Bbb I}$, the algebra contains the following
elements: 
\[
\begin{tabular}{ll}
Grade 0: & ${\Bbb I}$, $\theta \bar \theta $, $\theta \theta \theta $, $\bar
\theta \bar \theta \bar \theta $ \\ 
Grade 1: & $\theta $, $\bar \theta \bar \theta $ \\ 
Grade 2: & $\bar \theta $, $\theta \theta $%
\end{tabular}
\]
and its dimension is 
\[
D=1+2N+3N^2+2\frac{N^3-N}3=\frac{3+4N+9N^2+2N^3}3 
\]

One can note that the grade 0 elements recall formally the only observable
combinations of {\em quark fields} in chromodynamics based on the $SU(3)$
symmetry, that is the mesons which are the combinations $q\bar q$ of one
quark and one antiquark, and the hadrons which are the combinations $qqq$ or 
$\bar q\bar q\bar q$ of three quarks or three antiquarks.

From now on, we shall denote the grade of an object $X$ by $\partial X$.

\section{${\Bbb Z}_3$-matrices}

We define the ${\Bbb Z}_3$-matrices in analogy with the supermatrices, which
form a ${\Bbb Z}_2$-graded matrix algebra and whose entries are elements of
a Grassmann algebra.

First, we define a ${\Bbb Z}_3$-graded complex matrix algebra by dividing
the algebra ${\cal A}$ of $3\times 3$ block matrices into three parts ${\cal %
A}={\cal A}_0\oplus {\cal A}_1\oplus {\cal A}_2$. A matrix is an element of $%
{\cal A}_0$ (resp. ${\cal A}_1$, ${\cal A}_2$) and is of grade 0 (resp. 1,
2) if it has the form shown below: 
\[
\left( 
\begin{array}{lll}
A & 0 & 0 \\ 
0 & B & 0 \\ 
0 & 0 & C
\end{array}
\right) \mbox{, resp. }\left( 
\begin{array}{lll}
0 & A & 0 \\ 
0 & 0 & B \\ 
C & 0 & 0
\end{array}
\right) \mbox{, }\left( 
\begin{array}{lll}
0 & 0 & A \\ 
B & 0 & 0 \\ 
0 & C & 0
\end{array}
\right) \mbox{.} 
\]

This gives a ${\Bbb Z}_3$-graded structure to the algebra of matrices, in
the sense that the grade of the product of two matrices is the sum of the
grades of these matrices modulo 3.

We can then tensorize this graded algebra with our ${\Bbb Z}_3$-graded
Grassmann algebra, the grade of a ${\Bbb Z}_3$-matrix being the sum {\em %
modulo 3} of the grade of the matrix and of the grade of the Grassmann
element. So a ${\Bbb Z}_3$-matrix is of grade 0 (resp. 1, 2) if its blocks
contain only Grassmann elements with the respective grades:

\[
\stackunder{\mbox{grade 0}}{\left( 
\begin{array}{lll}
0 & 2 & 1 \\ 
1 & 0 & 2 \\ 
2 & 1 & 0
\end{array}
\right) }\mbox{, }\stackunder{\mbox{grade 1}}{\left( 
\begin{array}{lll}
1 & 0 & 2 \\ 
2 & 1 & 0 \\ 
0 & 2 & 1
\end{array}
\right) }\mbox{, }\stackunder{\mbox{grade 2}}{\left( 
\begin{array}{lll}
2 & 1 & 0 \\ 
0 & 2 & 1 \\ 
1 & 0 & 2
\end{array}
\right) } 
\]

It is easy to verify that the grade of the product of two ${\Bbb Z}_3$%
-matrices is the sum of their grades modulo 3.

The algebra of ${\Bbb Z}_3$-matrices contains a neutral element, ${\cal I}$,
the ${\Bbb Z}_3$-matrix whose only non-vanishing elements are the unit of
the Grassmann algebra occupying the main diagonal. The existence of this
element enables us to define the invertibility of a ${\Bbb Z}_3$-matrix.

\begin{theorem}
\begin{description}
\item[(a)]  A block of a ${\Bbb Z}_3$-matrix is invertible if and only if
the block matrix formed with the coefficients of ${\Bbb I}$ in the
development of the elements of the block over the grade 0 elements of the
Grassmann algebra (${\Bbb I}$, $\theta \bar \theta $, $\theta \theta \theta $%
, $\bar \theta \bar \theta \bar \theta $) is invertible.

\item[(b)]  \label{inver}A ${\Bbb Z}_3$-matrix is invertible if and only if
its grade 0 blocks are invertible. The inverse of a grade 1 (resp. 2, 0) $%
{\Bbb Z}_3$-matrix is a grade 2 (resp. 1, 0) ${\Bbb Z}_3$-matrix.

\item[(c)]  The product of two invertible ${\Bbb Z}_3$-matrices is an
invertible ${\Bbb Z}_3$-matrix.
\end{description}
\end{theorem}

\proof

\begin{enumerate}
\item[(a)]  Let us decompose a ${\Bbb Z}_3$-matrix $M$ into its complex
component matrices along the elements of the Grassmann algebra: $M=M_{\o} 
{\Bbb I}+M_\mu \theta _\mu +\bar M_\mu \bar \theta _\mu +M_{\mu \nu }\theta
_\mu \theta _\nu +\bar M_{\mu \nu }\bar \theta _\mu \bar \theta _\nu +%
\widetilde{M}_{\mu \nu }\theta _\mu \bar \theta _\nu +M_{\mu \nu \eta
}\theta _\mu \theta _\nu \theta _\eta +\bar M_{\mu \nu \eta }\bar \theta
_\mu \bar \theta _\nu \bar \theta _\eta $. We also consider another ${\Bbb Z}%
_3$-matrix $N$ that we decompose in the same way. Then it is easy to see
that the component of $MN$ along ${\Bbb I}$ is $M_{\o} N_{\o} $ (no product
of two generators different from ${\Bbb I}$ can give something proportional
to ${\Bbb I}$) so that for $M$ to be invertible, its component $M_{\o} $
must be invertible. Conversely, if $M_{\o} $ is invertible, let us put $%
N_{\o} =(M_{\o} )^{-1}$. Then the product $MN$ differs from ${\cal I}$ only
in terms of degree 2 and higher. However, we can choose $N_\mu =-(M_{\o}
)^{-1}M_\mu (M_{\o} )^{-1}$ and $\bar N_\mu =-(M_{\o} )^{-1}\bar M_\mu
(M_{\o} )^{-1}$. This way, $MN $ differs from ${\cal I}$ only in terms of
degree 3 and higher. Choosing the higher-level components of $N$ in the same
way, it is easy to expel these terms to higher degrees, until we have
actually constructed the inverse of $M $, because of the finite number of
terms in the development and of their cubic nilpotence.

\item[(b)]  For this part of the theorem, we can use the proof of part (a)
by noting that the component $M_{\o} $ of a matrix $M$ has non-zero
coefficients only in positions corresponding to grade 0 blocks. ${\cal I}$
being of grade 0, it is obvious that the sum of the grades of a ${\Bbb Z}_3$%
-matrix and of its inverse should be 0 modulo 3.

\item[(c)]  The component along ${\Bbb I}$ of the product of two invertible
matrices is the product of their components along ${\Bbb I}$, which are
invertible matrices by virtue of (b) and (a). These components being
ordinary matrices, their product is an invertible matrix. We use (b) once
more to conclude.\endproof
\end{enumerate}

The left (resp. right) product of a supermatrix by an element $\lambda $ of
the ordinary Grassmann algebra is defined as its left (resp. right)
multiplication by the supermatrix 
\[
\left( 
\begin{array}{ll}
\lambda &  \\ 
& (-1)^{\partial \lambda }\lambda
\end{array}
\right) 
\]

Generalizing this idea, we define the left (resp. right) product of a ${\Bbb %
Z}_3$-matrix by an element $\lambda $ of the Grassmann algebra as its left
(resp. right) multiplication by the following diagonal ${\Bbb Z}_3$-matrix: 
\[
\left( 
\begin{array}{lll}
\lambda &  &  \\ 
& j^{2\partial \lambda }\lambda &  \\ 
&  & j^{\partial \lambda }\lambda
\end{array}
\right) 
\]

Note that in general, $\lambda M\neq M\lambda $, but we have $\lambda
(MN)=(\lambda M)N$, $(M\lambda )N=M(\lambda N)$, $M(N\lambda )=(MN)\lambda $
and $\lambda (M\mu )=(\lambda M)\mu $ which give our algebra the structure
of a bimodule with respect to the Grassmann algebra.

\begin{definition}
Let us define a ${\Bbb Z}_3$-matrix $M$ with the following block structure:

\[
M=\left( 
\begin{array}{lll}
A & B & C \\ 
D & E & F \\ 
G & H & I
\end{array}
\right) 
\]

Its {\em ${\Bbb Z}_3$-trace }is defined by:{\em \ }$\limfunc{tr}\nolimits_{%
{\Bbb Z}_3}(M)=\limfunc{tr}(A)+j^{2(1-\partial M)}\limfunc{tr}%
(E)+j^{(1-\partial M)}\limfunc{tr}(I)$
\end{definition}

that is:

\begin{itemize}
\item  If $M$ is of grade 0, then $\limfunc{tr}\nolimits_{{\Bbb Z}_3}(M)=%
\limfunc{tr}(A)+j^2.\limfunc{tr}(E)+j.\limfunc{tr}(I)$

\item  If $M$ is of grade 1, then $\limfunc{tr}\nolimits_{{\Bbb Z}_3}(M)=%
\limfunc{tr}(A)+\limfunc{tr}(E)+\limfunc{tr}(I)$

\item  If $M$ is of grade 2, then $\limfunc{tr}\nolimits_{{\Bbb Z}_3}(M)=%
\limfunc{tr}(A)+j\limfunc{tr}(E)+j^2\limfunc{tr}(I)$
\end{itemize}

The supertrace of a supermatrix $M=\left( 
\begin{array}{ll}
A & B \\ 
C & D
\end{array}
\right) $was defined by $\limfunc{str}M=\limfunc{tr}(A)+(-1)^{1-\partial M}%
\limfunc{tr}(D)$.

The following theorem can be easily proved:

\begin{theorem}
\begin{description}
\item[(a)]  If $M$ and $N$ are of the same grade, 
\[
\limfunc{tr}\nolimits_{{\Bbb Z}_3}(M+N)=\limfunc{tr}\nolimits_{{\Bbb Z}%
_3}(M)+\limfunc{tr}\nolimits_{{\Bbb Z}_3}(N) 
\]

\item[(b)]  $\limfunc{tr}\nolimits_{{\Bbb Z}_3}(\lambda M)=\lambda \limfunc{%
tr}\nolimits_{{\Bbb Z}_3}(M)$ and $\limfunc{tr}\nolimits_{{\Bbb Z}%
_3}(M\lambda )=\limfunc{tr}\nolimits_{{\Bbb Z}_3}(M)\lambda $

\item[(c)] 
\begin{itemize}
\item  If $M$ and $N$ are of grade $0$ then $\limfunc{tr}\nolimits_{{\Bbb Z}%
_3}(MN)=\limfunc{tr}\nolimits_{{\Bbb Z}_3}(NM)$

\item  If $M$ is of grade $1$ and $N$ of grade $2$, then $\limfunc{tr}%
\nolimits_{{\Bbb Z}_3}(MN)=j\limfunc{tr}\nolimits_{{\Bbb Z}_3}(NM)$
\end{itemize}
\end{description}
\end{theorem}

\begin{corollary}
\begin{itemize}
\item  If $M,N$ and $P$ are of grade $1$ then $\limfunc{tr}\nolimits_{{\Bbb Z%
}_3}(MNP)=j\limfunc{tr}\nolimits_{{\Bbb Z}_3}(NPM)$

\item  If $M,N$ and $P$ are of grade $2$ then $\limfunc{tr}\nolimits_{{\Bbb Z%
}_3}(MNP)=j^2\limfunc{tr}\nolimits_{{\Bbb Z}_3}(NPM)$
\end{itemize}
\end{corollary}

The proofs are straightforward, and the only non trivial observation
concerns the fact that $\limfunc{tr}(MN)=\limfunc{tr}(NM)$ is not always
true if $M$ and $N$'s coefficients are not numbers.

Finally, we can define the generalization of the superdeterminant this way:

\begin{definition}
${\Bbb Z}_3${\bf -determinant:}{\em \ }If $M$ is a {\em grade 0} invertible $%
{\Bbb Z}_3$-matrix, then its {\em ${\Bbb Z}_3$-determinant }is 
\[
\begin{array}{r}
\det_{{\Bbb Z}_3}(M)=\det
(A-CI^{-1}G-(B-CI^{-1}H)(E-FI^{-1}H)^{-1}(D-FI^{-1}G))\times \\ 
\times(\det (E-FI^{-1}H))^{j^2}(\det I)^j
\end{array}
\]
\end{definition}

\begin{theorem}
If $M$ and $N$ are two invertible grade 0 ${\Bbb Z}_3$-matrices, 
\[
\det\nolimits_{{\Bbb Z}_3}(MN)=(\det\nolimits_{{\Bbb Z}_3}M)(\det\nolimits_{%
{\Bbb Z}_3}N) 
\]
\label{detmul}
\end{theorem}

\proof Any grade 0 invertible ${\Bbb Z}_3$-matrix can be
decomposed into the product of three ${\Bbb Z}_3$-matrices: $M=M_1M_0M_2$
with 
\begin{equation}
\begin{array}{l}
M_1=\left( {\footnotesize \renewcommand{\arraystretch}{0.5}%
\begin{tabular}{ccc}
$1$ & $(B-CI^{-1}H)(E-FI^{-1}H)^{-1}$ & $CI^{-1}$ \\ 
$0$ & $1$ & $FI^{-1}$ \\ 
$0$ & $0$ & $1$%
\end{tabular}
}\right) \\ 
M_0=\left( {\footnotesize \renewcommand{\arraystretch}{0.5}%
\begin{tabular}{ccc}
$A-CI^{-1}G-(B-CI^{-1}H)(E-FI^{-1}H)^{-1}(D-FI^{-1}G)$ & $0$ & $0$ \\ 
$0$ & $E-FI^{-1}H$ & $0$ \\ 
$0$ & $0$ & $I$%
\end{tabular}
}\right) \\ 
M_2=\left( {\footnotesize \renewcommand{\arraystretch}{0.5}%
\begin{tabular}{ccc}
$1$ & $0$ & $0$ \\ 
$(D-FI^{-1}G)(E-FI^{-1}H)^{-1}$ & $1$ & $0$ \\ 
$I^{-1}G$ & $I^{-1}H$ & $1$%
\end{tabular}
}\right)
\end{array}
\label{decompo}
\end{equation}
It is very easy to show that the theorem is true for any $M$ if $N$ is
block-diagonal or inferior-block-triangular with the identity in the
diagonal blocks. It is also easy to generalize to an arbitrary $N$ in the
case where $M$ is block-diagonal or superior-block-triangular with the
identity in the diagonal blocks. Therefore, we have: 
\[
\ 
\begin{array}{r}
\det_{{\Bbb Z}_3}(MN)=\det_{{\Bbb Z}_3}(M_1)\det_{{\Bbb Z}_3}(M_0)\det_{%
{\Bbb Z}_3}(M_2N_1)\det_{{\Bbb Z}_3}(N_0)\det_{{\Bbb Z}_3}(N_2)= \\ 
=\det_{{\Bbb Z}_3}(M)\det_{{\Bbb Z}_3}(M_2N_1)\det_{{\Bbb Z}_3}(N)
\end{array}
\]
and the theorem remains to be proved only for $M$ of the form $\left( %
\renewcommand{\arraystretch}{0.5}
\begin{tabular}{ccc}
$1$ & $0$ & $0$ \\ 
$A$ & $1$ & $0$ \\ 
$B$ & $C$ & $1$
\end{tabular}
\right) $ and $N$ of the form $\left( \renewcommand{\arraystretch}{0.5}%
\begin{tabular}{ccc}
$1$ & $D$ & $E$ \\ 
$0$ & $1$ & $F$ \\ 
$0$ & $0$ & $1$
\end{tabular}
\right) $. One can note that $\left( \renewcommand{\arraystretch}{0.5}%
\begin{tabular}{ccc}
$1$ & $0$ & $B$ \\ 
$0$ & $1$ & $C$ \\ 
$0$ & $0$ & $1$
\end{tabular}
\right) \left( \renewcommand{\arraystretch}{0.5}
\begin{tabular}{ccc}
$1$ & $0$ & $0$ \\ 
$0$ & $1$ & $C^{\prime }$ \\ 
$0$ & $0$ & $1$
\end{tabular}
\right) =\left( \renewcommand{\arraystretch}{0.5}
\begin{tabular}{ccc}
$1$ & $0$ & $B$ \\ 
$0$ & $1$ & $C+C^{\prime }$ \\ 
$0$ & $0$ & $1$
\end{tabular}
\right) $, $\left( \renewcommand{\arraystretch}{0.5}
\begin{tabular}{ccc}
$1$ & $A$ & $B$ \\ 
$0$ & $1$ & $C$ \\ 
$0$ & $0$ & $1$
\end{tabular}
\right) \left( \renewcommand{\arraystretch}{0.5}
\begin{tabular}{ccc}
$1$ & $A^{\prime }$ & $0$ \\ 
$0$ & $1$ & $0$ \\ 
$0$ & $0$ & $1$
\end{tabular}
\right) =\left( \renewcommand{\arraystretch}{0.5}
\begin{tabular}{ccc}
$1$ & $A+A^{\prime }$ & $B$ \\ 
$0$ & $1$ & $C$ \\ 
$0$ & $0$ & $1$
\end{tabular}
\right) $ and $\left( \renewcommand{\arraystretch}{0.5}
\begin{tabular}{ccc}
$1$ & $A$ & $B$ \\ 
$0$ & $1$ & $C$ \\ 
$0$ & $0$ & $1$
\end{tabular}
\right) \left( \renewcommand{\arraystretch}{0.5}
\begin{tabular}{ccc}
$1$ & $0$ & $B^{\prime }$ \\ 
$0$ & $1$ & $0$ \\ 
$0$ & $0$ & $1$
\end{tabular}
\right) =\left( \renewcommand{\arraystretch}{0.5}
\begin{tabular}{ccc}
$1$ & $A$ & $B+B^{\prime }$ \\ 
$0$ & $1$ & $C$ \\ 
$0$ & $0$ & $1$
\end{tabular}
\right) $ so that we can construct the matrix $N_1$, element by element,
starting from the matrix that contains only its diagonal blocks, using left
multiplication by a series of matrices that contain only one non-zero
element out of the diagonal blocks.

We show the theorem for $M$ of the form of $M_2$ and $N$ containing only one
element out of the diagonal blocks, using the fact that if $\partial X\neq 0$%
, then $(1+X)^\alpha =1+\alpha X+\frac{\alpha (\alpha -1)}2X^2$. If $%
N_1=N_1^0N_1^1$ we can decompose $M_2N_1^0$ in the form \ref{decompo}. Then,

\[
\ 
\begin{array}{r}
\det_{{\Bbb Z}_3}(M_2N_1)=\det_{{\Bbb Z}%
_3}((M_2N_1^0)_1(M_2N_1^0)_0(M_2N_1^0)_2.N_1^1)= \\ 
=\det_{{\Bbb Z}_3}((M_2N_1^0)_0)\det_{{\Bbb Z}_3}((M_2N_1^0)_2.N_1^1)=\det_{%
{\Bbb Z}_3}((M_2N_1^0)_0)
\end{array}
\]
if $N_1^1$ contains only one element out of its diagonal blocks. But $\det_{%
{\Bbb Z}_3}((M_2N_1^0)_0)=\det_{{\Bbb Z}_3}(M_2N_1^0)$. We can perform the
same operation, until only the product of $M_2$ and the matrix formed with
the diagonal blocks of $N_1$ (which are the identity) remains. Thus $\det_{%
{\Bbb Z}_3}(M_2N_1)=1$.\endproof

\begin{theorem}
If $M$ is a grade 0 invertible ${\Bbb Z}_3$-matrix, its ${\Bbb Z}_3$%
-determinant can also be expressed in the following five alternative ways:

\[
\ {\ 
\begin{array}[t]{l}
\det_{{\Bbb Z}_3}(M)=\det (A-CI^{-1}G)\times \\ 
\times(\det (E-FI^{-1}H-(D-FI^{-1}G)(A-CI^{-1}G)^{-1}(B-CI^{-1}H)))^{j^2}(\det
I)^j
\end{array}
} 
\]

\[
\ {\ 
\begin{array}[t]{l}
\det_{{\Bbb Z}_3}(M)=(\det A).(\det
(E-DA^{-1}B-(F-DA^{-1}C)(I-GA^{-1}C)^{-1}(H-GA^{-1}B)))^{j^2}\times \\ 
\times(\det (I-GA^{-1}C))^j
\end{array}
} 
\]

\[
\ {\ 
\begin{array}[t]{l}
\det_{{\Bbb Z}_3}(M)=(\det A)(\det (E-DA^{-1}B))^{j^2}\times \\ 
\times(\det (I-GA^{-1}C-(H-GA^{-1}B)(E-DA^{-1}B)^{-1}(F-DA^{-1}C)))^j
\end{array}
} 
\]

\[
\ {\ 
\begin{array}[t]{l}
\det_{{\Bbb Z}_3}(M)=(\det
(A-BE^{-1}D-(C-BE^{-1}F)(I-HE^{-1}F)^{-1}(G-HE^{-1}D)))\times \\ 
\times(\det E)^{j^2}(\det (I-HE^{-1}F))^j
\end{array}
} 
\]

\[
\ {\ 
\begin{array}[t]{l}
\det_{{\Bbb Z}_3}(M)=(\det (A-BE^{-1}D)(\det E)^{j^2}\times \\ 
\times(\det (I-HE^{-1}F-(G-HE^{-1}D)(A-BE^{-1}D)^{-1}(C-BE^{-1}F)))^j
\end{array}
} 
\]
\end{theorem}

\proof We use the following decompositions of $M$:

\[
\left( {\tiny \renewcommand{\arraystretch}{0.65}
\begin{tabular}{lll}
$A-CI^{-1}G$ & $0$ & $C$ \\ 
$D-FI^{-1}G$ & $E-FI^{-1}H-(D-FI^{-1}G)(A-CI^{-1}G)^{-1}(B-CI^{-1}H)$ & $F$
\\ 
$0$ & $0$ & $I$%
\end{tabular}
} \right) \left( {\tiny \renewcommand{\arraystretch}{0.65}
\begin{tabular}{lll}
$1$ & $(A-CI^{-1}G)^{-1}(B-CI^{-1}H)$ & $0$ \\ 
$0$ & $1$ & $0$ \\ 
$I^{-1}G$ & $I^{-1}H$ & $1$%
\end{tabular}
} \right) 
\]

\[
\left( {\tiny \renewcommand{\arraystretch}{0.65}
\begin{tabular}{lll}
$1$ & $0$ & $0$ \\ 
$DA^{-1}$ & $1$ & $(F-DA^{-1}C)(I-GA^{-1}C)^{-1}$ \\ 
$GA^{-1}$ & $0$ & $1$%
\end{tabular}
} \right) \left( {\tiny \renewcommand{\arraystretch}{0.65}
\begin{tabular}{lll}
$A$ & $B$ & $C$ \\ 
$0$ & $E-DA^{-1}B-(F-DA^{-1}C)(I-GA^{-1}C)^{-1}(H-GA^{-1}B)$ & $0$ \\ 
$0$ & $H-GA^{-1}B$ & $I-GA^{-1}C$%
\end{tabular}
} \right) 
\]

\[
\left( {\tiny \renewcommand{\arraystretch}{0.65}
\begin{tabular}{lll}
$1$ & $0$ & $0$ \\ 
$DA^{-1}$ & $1$ & $0$ \\ 
$GA^{-1}$ & $(H-GA^{-1}B)(E-DA^{-1}B)^{-1}$ & $1$%
\end{tabular}
} \right) \left( {\tiny \renewcommand{\arraystretch}{0.65}
\begin{tabular}{lll}
$A$ & $B$ & $C$ \\ 
$0$ & $E-DA^{-1}B$ & $F-DA^{-1}C$ \\ 
$0$ & $0$ & $I-GA^{-1}C-(H-GA^{-1}B)(E-DA^{-1}B)^{-1}(F-DA^{-1}C)$%
\end{tabular}
} \right) 
\]

\[
\left( {\tiny \renewcommand{\arraystretch}{0.65}
\begin{tabular}{lll}
$1$ & $BE^{-1}$ & $(C-BE^{-1}F)(I-HE^{-1}F)^{-1}$ \\ 
$0$ & $1$ & $0$ \\ 
$0$ & $HE^{-1}$ & $1$%
\end{tabular}
} \right) \left( {\tiny \renewcommand{\arraystretch}{0.65}
\begin{tabular}{lll}
$A-BE^{-1}D-(C-BE^{-1}F)(I-HE^{-1}F)^{-1}(G-HE^{-1}D)$ & $0$ & $0$ \\ 
$D$ & $E$ & $F$ \\ 
$0$ & $0$ & $I-HE^{-1}F$%
\end{tabular}
} \right) 
\]

\[
\ \left( {\tiny \renewcommand{\arraystretch}{0.65}
\begin{tabular}{lll}
$A-BE^{-1}D$ & $B$ & $0$ \\ 
$0$ & $E$ & $0$ \\ 
$G-HE^{-1}D$ & $H$ & $I-HE^{-1}F-(G-HE^{-1}D)(A-BE^{-1}D)^{-1}(C-BE^{-1}F)$%
\end{tabular}
}\right) \left( {\tiny \renewcommand{\arraystretch}{0.65}
\begin{tabular}{lll}
$1$ & $0$ & $(A-BE^{-1}D)^{-1}(C-BE^{-1}F)$ \\ 
$E^{-1}D$ & $1$ & $E^{-1}F$ \\ 
$0$ & $0$ & $1$%
\end{tabular}
}\right) 
\]
and make use of the theorem \ref{detmul}.\endproof

In the case of the superdeterminant, the following two expressions are
equivalent: 
\[
\limfunc{sdet}\left( 
\begin{array}{cc}
A & B \\ 
C & D
\end{array}
\right) =\det (A-BD^{-1}C).(\det D)^{-1}=\det A.(\det (D-CA^{-1}B))^{-1} 
\]
This restores the symmetry between the two diagonal elements on the one
hand, and between the two off diagonal elements on the other hand. Here we
have six expressions, restoring the $S_3$ symmetry between the three grade
0, grade 1 and grade 2 elements of the ${\Bbb Z}_3$-matrix which is broken
in any expression separately chosen.

\begin{theorem}
If $M$ is a grade 0 invertible $3\times 3$ ${\Bbb Z}_3$-matrix, then 
\[
\det\nolimits_{{\Bbb Z}_3}(\exp M)=\exp (\limfunc{tr}\nolimits_{{\Bbb Z}_3}M)
\]
\end{theorem}

\proof Let $M$ be the matrix: 
\[
M=\left( 
\begin{array}{ccc}
a & \bar A & A \\ 
B & b & \bar B \\ 
\bar C & C & c
\end{array}
\right) 
\]
where $a,b,c$ are three grade 0 invertible elements of the Grassmann algebra
(they must be invertible in order for $M$ to be invertible, see theorem \ref
{inver}), $A,B,C$ are three grade 1 elements of the Grassmann algebra, and $%
\bar A,\bar B,\bar C$ are three grade 2 elements of the Grassmann algebra.
Assuming that $a,b,c$ are distinct, straightforward calculus give the
following expression for the exponential of the matrix $M$: 
\[
\exp M=\left( 
\begin{array}{l}
e^a+f(a,b)\bar AB+f(a,c)A\bar C+g(a,b,c)(\bar A\bar B\bar C+j.CBA) \\ 
h(a,b)B+l(a,b,c)\bar B\bar C \\ 
h(a,c)\bar C+l(a,b,c)CB
\end{array}
\right.
\]
\[ 
\begin{array}{c}
h(a,b)\bar A+l(a,b,c)AC \\ 
e^b+j^2f(b,c)C\bar B+j.f(b,a)\bar AB+j.g(b,c,a)(\bar A\bar B\bar C+j.CBA) \\ 
h(b,c)C+l(a,b,c)\bar C\bar A
\end{array}
\]
\[
\left. 
\begin{array}{r}
h(a,c)A+l(a,b,c)\bar A\bar B \\ 
h(b,c)\bar B+l(a,b,c)BA \\ 
e^c+j^2.f(c,a)A\bar C+f(c,b)C\bar B+j^2.g(c,a,b)(\bar A\bar B\bar C+j.CBA)
\end{array}
\right) 
\]
where 
\[
f(x,y)=\frac{(x-y-1)e^x+e^y}{(x-y)^2}
\]
\[
\begin{array}{r}
g(x,y,z)=\left[ \frac 1{(x-y)(x-z)}-\frac 1{(x-y)(x-z)^2}-\frac
1{(x-z)(x-y)^2}\right] e^x+ \\ 
+\frac{e^y}{(y-z)(y-x)^2}+\frac{e^z}{(z-y)(z-x)^2}
\end{array}
\]
\[
h(x,y)=\frac{e^x-e^y}{x-y}
\]
\[
l(x,y,z)=\frac{e^x}{(x-y)(x-z)}+\frac{e^y}{(y-z)(y-x)}+\frac{e^z}{(z-x)(z-y)}
\]
From this we compute 
\[
\begin{array}{r}
\det\nolimits_{{\Bbb Z}_3}(\exp M)=e^{a+j^2b+jc}\left[ {\Bbb I}%
+(e^{-a}f(a,b)+e^{-b}f(b,a)-e^{-(a+b)}(h(a,b))^2)\bar AB+\right.  \\ 
+(e^{-a}f(a,c)+e^{-c}f(c,a)-e^{-(a+c)}(h(a,c))^2)A\bar C+ \\ 
+j(e^{-c}f(c,b)+e^{-b}f(b,c)-e^{-(b+c)}(h(b,c))^2)C\bar B+ \\ 
+(e^{-a}g(a,b,c)+e^{-b}g(b,c,a)+e^{-c}g(c,a,b)+e^{-(a+b+c)}h(a,b)h(b,c)h(c,a)-
\\ 
\left. -(e^{-(a+b)}h(a,b)+e^{-(a+c)}h(a,c)+e^{-(b+c)}h(b,c))l(a,b,c))(\bar
A\bar B\bar C+j.CBA)\right] 
\end{array}
\]
and it is then easy to verify that all terms
inside the brackets vanish, except for ${\Bbb I}$. The cases where $a=b$, $%
b=c$ or $c=a$ are just limits of the previous case.
\endproof

These constructions will be used in a forthcoming paper concerning the
construction of a gauge theory based on a ${\Bbb Z}_3$-graded
non-commutative geometry model similar to the one used by Coquereaux {\em et
al.}\cite{coquereaux1991}, using instead Kerner's differential whose cube is
zero, whereas its square is not\cite{kerner1995}.

\section{Aknowledgments}
I wish to thank Profs. Kerner and Abramov for fruitful discussions.
This work has been financed by a grant from the French {\em Minist\`ere de l'Enseignement
Sup\'erieur et de la Recherche}.

\end{document}